\documentclass[a4paper, 12pt]{article}
\usepackage[english]{babel}
\usepackage[a4paper, inner=2cm, outer=2cm, top=3cm, bottom=3cm]{geometry}
\usepackage[dvipsnames]{xcolor}
\usepackage{mathtools}
\usepackage{pstricks}
\usepackage{caption}
\usepackage{graphicx}
\usepackage{amsmath}
\usepackage{amssymb}
\usepackage{mathcomp}
\usepackage{cancel}
\usepackage{textcomp}
\usepackage{enumitem}
\usepackage{physics}
\usepackage{romannum}
\usepackage[doublespacing]{setspace}
\usepackage{subcaption}
\usepackage{hyperref}
\usepackage{xcolor}
\usepackage{booktabs}
\usepackage{cite}

\definecolor{DarkBlue}{RGB}{0,38,77}

\hypersetup{
  colorlinks   = true,
  urlcolor     = Blue, 
  linkcolor    = Black,
  citecolor    = DarkBlue
}
\renewcommand{\thesection}{\Roman{section}}

\title{On the Gravitational Hysteresis in the Kinetic Theory}
\author{Raihaneh Moti$^1$ and Ali Shojai$^2$\\
\textit{{\small $^1$ School of Astronomy, Institute for Research in Fundamental Sciences (IPM), Tehran, Iran}} \\ \textit{{\small $^2$ Department of Physics, University of Tehran, North Karegar St., Tehran, Iran}}}
\date{}
\begin{document}
\maketitle
\pagenumbering{arabic}
\begin{abstract}
General theory of relativity is non--linear in nature and therefore can result in hysteresis--like effects and cause systems to remember the footprint of the gravitational field. Here we have investigated this effect using the Kinetic theory in curved spacetime. It is shown that the entropy rate experiences this hysteresis behavior. The effect is then considered for some special spacetimes, including Schwarzschild black hole, Friedmann--Lema\^itre--Robertson--Walker cosmological solution, and the flat Minkowski spacetime perturbed by a gravitational wave pulse. It is shown that there is some hysteresis effect for the entropy rate.
\end{abstract}
\section{Introduction}
The intrinsic non--linearity of gravitational field, is a fundamental feature of the general theory of relativity. Unlike electrodynamics, this property is classical and does not stem from quantum mechanical effects.
Therefore, the superposition principle is no longer applicable at the classical level, and there is an observable gravitational hysteresis effect even in the weak field limit. 
This is what sometimes referred to as the memory effect.

One definition for this property of gravity is based on the persistent changes in the polarization components (at least one of them) of the gravitational wave
\begin{equation}
\Delta h_{+,\times}^{\text{mem}} = \lim_{t\to +\infty} h_{+,\times}(t) - \lim_{t\to -\infty} h_{+,\times}(t)
\end{equation}
where $t$ is the time at the observer location \cite{Braginsky:1987kwo, Favata:2009ii}.

Sometimes, this effect (misleadingly) is categorized as linear or non--linear.  The former is defined as the change in the time derivatives of the multipole moments of the sources at nearly zero frequency \cite{Zeldovich:1974gvh, Smarr:1977fy, Kovacs:1978eu, Braginsky:1985vlg}. It also appears in systems undergoing a kick or ejecting particles in an anisotropic way \cite{Favata:2008ti,Favata:2008yd}. On the other hand, the nonlinear one, also known as Christodoulo memory, originates from the unbound gravitons radiated by the system \cite{Christodoulou:1991cr, Blanchet:1992br}. Since the emitted radiation results in a loss of gravitational energy, the multipole moments of the source would change, and since the gravitational wave is associated with these moments, it is also affected by the graviton radiation. \cite{Thorne:1992sdb, Favata:2009ii, Favata:2010zu}.

A gravitational wave pulse in the linear theory has an oscillating amplitude with its profile  starting from a small value, growing to a peak, and then diminishing to zero eventually. But a realistic gravitational wave does not follow this pattern because of the presence of a memory sector, which causes a permanent deformation after the passage through an idealized detector (i.e. one that is truly freely falling). This means that we can have {\em hysteresis} effects. It may appear in various types such as displacement \cite{Zhang:2017geq}, kick \cite{Grishchuk:1989qa, Divakarla:2021xrd}, spin \cite{Barnich:2009se, Pasterski:2015tva, Nichols:2017rqr}, gyroscope memory \cite{Seraj:2021rxd, Seraj:2022qyt}, etc. Because of the fact that these deformations are footprints of the wave, they may potentially be used for the detection of the gravitational waves \cite{Moti:2023tey}.

It is clear that the presence of such hysteresis aspect, either in the presence of gravitational wave pulses or within any gravitational background, implies a preferred direction of time. This means that the entropy (and its rate) can be a good probe for the effect (like what is done in \cite{Moti:2023tey}).

To address this, here we consider a box of fluid from the kinetic theoretical point of view and use the Boltzmann equation in curved spacetime (with distribution function $f \equiv f(\vb{x}, \vb{v}, t) $) for investigating the evolution of the fluid. Usually the method of moments would ease the complexity of dealing with the macroscopic quantities. 
These moments provide us a deeper understanding of the physics of the system without requiring direct solutions to the Boltzmann equation. This is the approach we will follow in this paper.
In section \ref{SecII}, we will obtain the general effect of the gravitational field on the kinetic moments of such a fluid. Then in section \ref{SecIII}, some special cases are investigated.
 
\section{Kinetic moments in the curved spacetime}
\label{SecII}
It is known that the macroscopic description of a flow is best described by evolution equation of a few number of moments of the one--particle Boltzmann distribution function. In curved spacetime the non--linear nature of the gravitational theory can result in hysteresis--like behavior.  

To investigate such an effect, let us consider a box containing a fluid which can be described by the Boltzmann distribution function $f(x,p)$ and its moments. As it is reviewed in the \ref{App. A}, the distribution function satisfies the Boltzmann equation
\begin{equation}
p^{\mu} \pdv{f}{x^{\mu}} - \Gamma^{i}_{\mu\nu} p^{\mu} p^{\nu} \pdv{f}{p^{i}} = Q(f,F,\Omega,p)
\end{equation}
where $Q(f,F,\Omega,p)$ term contains collision effects in which $\Omega$ is the solid angle element characterizing binary collisions between the particles of the fluid and $F$ as defined in the \ref{App. A} is the invariant particle flux. 

It can be simply shown (see \ref{App. A}) that the three essential moments (particle number current, energy--momentum tensor and entropy current) defined as

\begin{equation}
    \begin{cases}
      & N^{\mu} = c \displaystyle\int p^{\mu} f \sqrt{g} \dfrac{\dd[3]p}{p_0}\\
      & T^{\mu\nu} = c \displaystyle\int p^{\mu} p^{\nu} f \sqrt{g} \dfrac{\dd[3]p}{p_0}\\
      & S^{\mu} = -ck_B  \displaystyle\int  p^{\mu} \ln{(\dfrac{f h^3}{e g_s})}    f \sqrt{g} \dfrac{\dd[3]{p}}{p_0}  
    \end{cases}
\label{EsMom}  
\end{equation}
obey the following relations

\begin{equation}
    \begin{cases}
      & N^{\mu}_{\ ;\mu} = 0 \\
      & T^{\mu\nu}_{\ \ ;\mu}  = 0\\
      & S^{\mu}_{\ ;\mu} \geq 0
    \end{cases}       
\end{equation}
where the $k_B$ is the Boltzmann constant.

\begin{figure}[h!]
\centering
\includegraphics[keepaspectratio=true, width=0.5\textwidth]{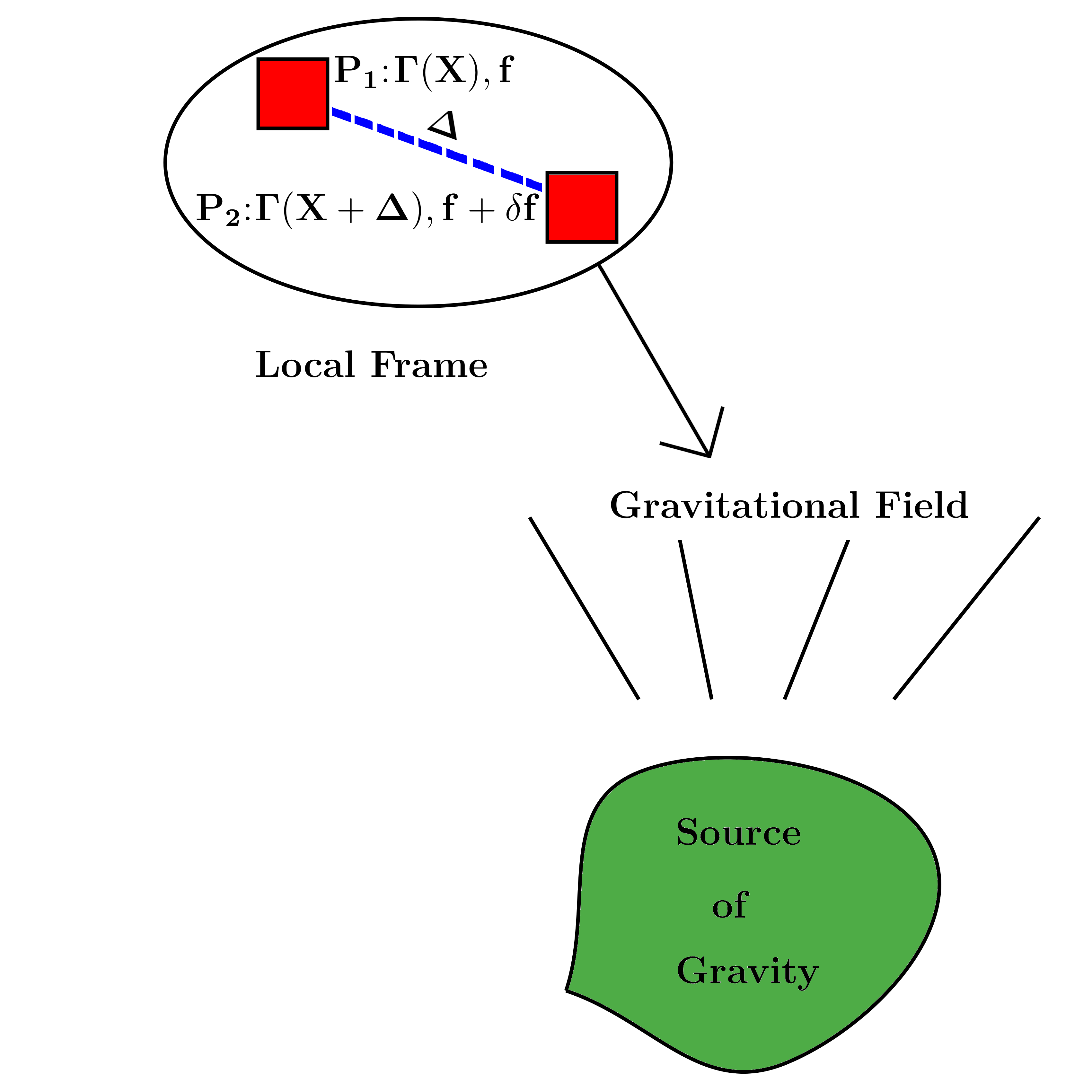}
\caption{A small box of fluid put in two neighboring places.}
 \label{Fig1}
\end{figure}

Now suppose that a small box of fluid (small compared to the characteristic length of the gravitational field) has been put either at point $P_1$ (at coordinates $X^\alpha$) or at point $P_2$ (at coordinates $X^\alpha+\Delta^\alpha$, where $\Delta^\alpha$ is some small displacement) as it is shown in figure \ref{Fig1}. 

It is clear that for the box at point $P_1$, we have
\begin{equation}
p^{\mu} \pdv{f}{x^{\mu}} - \Gamma^{i}_{\mu\nu}(X) p^{\mu} p^{\nu} \pdv{f}{p^{i}} = Q(f,F,\Omega,p) \ .
\end{equation}

Due to equations \eqref{LHS-App} and \eqref{RHS-App}, with a proper choice of the function $\psi \equiv \psi(x^{\mu},p^{\nu})$,  the evolution equation of any moment is
\begin{equation}
\left[ \int \psi p^{\mu} f \sqrt{g} \dfrac{\dd[3]{p}}{p_0}  \right]_{;\mu} = \int p^{\mu} f (\partial_{\mu} \psi) \sqrt{g} \dfrac{\dd[3]p}{p_0} - \int \pdv{\psi}{p_i} \Gamma^{i}_{\mu\nu}(X) p^{\mu} p^{\nu} f \sqrt{g} \dfrac{\dd[3]p}{p_0} + \text{\textit{RHS}}_1
\label{BEq.p1}
\end{equation}
where 
\begin{align}
\text{\textit{RHS}}_1 & = \int \psi Q (f,F,\Omega,p) \sqrt{g}  \dfrac{\dd[3]{p}}{p_0}  \nonumber \\
& = \int \psi \left(f' f'_* -  f f_* \right) F \sigma \dd{\Omega} \sqrt{g} \dfrac{\dd[3]{p_*}}{p_{*0}} \sqrt{g} \dfrac{\dd[3]{p}}{p_0} \nonumber \\
& = \dfrac{1}{4}  \int \left( \psi +\psi_* - \psi' - \psi'_*   \right) \left(f' f'_* -  f f_* \right) F \sigma \dd{\Omega} \sqrt{g} \dfrac{\dd[3]{p_*}}{p_{*0}} \sqrt{g} \dfrac{\dd[3]{p}}{p_0} \ .
\end{align}
The third equality in the equation is due to symmetry considerations.  By switching the roles of the colliding particles before and after the collision, the roles of their labels, specifically their momenta, also change in the integral in a symmetric way. This allows one to write the integrand as a quarter of the sum of the integrands for permutations of the particles (see for example \cite{Cercignani:2002}).

Alternatively, for the box at point $P_2$, we have (a bar over any quantity shows that it is related to the box at point 2)
\begin{equation}
p^{\mu} \pdv{\bar{f}}{x^{\mu}} - \bar{\Gamma}^{i}_{\mu\nu}(X+\Delta) p^{\mu} p^{\nu} \pdv{\bar{f}}{p^{i}} = \bar{Q} (\bar{f},\bar{F},\bar{\Omega},p)
\end{equation}
and, for the $\psi$-moments 
\begin{equation}
\left[ \int \psi p^{\mu} \bar{f} \sqrt{g} \dfrac{\dd[3]{p}}{p_0}  \right]_{;\mu} = \int p^{\mu} \bar{f} (\partial_{\mu} \psi) \sqrt{g} \dfrac{\dd[3]p}{p_0} - \int \pdv{\psi}{p_i} \bar{\Gamma}^{i}_{\mu\nu}(X + \Delta) p^{\mu} p^{\nu} \bar{f} \sqrt{g} \dfrac{\dd[3]p}{p_0} + \text{\textit{RHS}}_2
\label{BEq.p2}
\end{equation}
where
\begin{align}
\text{\textit{RHS}}_2 & = \int \bar{\psi}\bar{Q}(\bar{f},\bar{F},\bar{\Omega},p) \sqrt{g} \dfrac{\dd[3]{p}}{p_0}  \nonumber \\ &= \dfrac{1}{4}  \int \left( \bar{\psi} + \bar{\psi}_* - \bar{\psi}' - \bar{\psi}'_*   \right) \left(\bar{f}' \bar{f}'_* -  \bar{f} \bar{f}_* \right) \bar{F} \bar{\sigma} \dd{\bar{\Omega}} \sqrt{g} \dfrac{\dd[3]{p_*}}{p_{*0}} \sqrt{g} \dfrac{\dd[3]{p}}{p_0} \ .
\end{align}

To see what happens for a box of fluid when transported from point $P_1$ to point $P_2$, as stated previously we consider that the size of the box is less than the characteristic length of gravitational field (i.e. the inverse of the square root of the Riemann tensor). Comparison between these two points can be done via subtraction of equation \eqref{BEq.p1} from equation \eqref{BEq.p2}
\begin{align}
\left[ \int \psi p^{\mu} \bar{f} \sqrt{g} \dfrac{\dd[3]{p}}{p_0}  \right]_{;\mu} & - \left[ \int \psi p^{\mu} f \sqrt{g} \dfrac{\dd[3]{p}}{p_0}  \right]_{;\mu}
 = - \int p^{\mu} \left( \bar{f} \partial_{\mu} \bar{\psi} - f \partial_{\mu} \psi \right) \sqrt{g} \dfrac{\dd[3]p}{p_0} \nonumber \\
& \quad + \int \left( \pdv{\bar{\psi}}{p_i} \bar{\Gamma}^{i}_{\mu\nu}(X + \Delta) \bar{f} - \pdv{\psi}{p_i} \Gamma^{i}_{\mu\nu}(X) f  \right) p^{\mu} p^{\nu} \sqrt{g} \dfrac{\dd[3]{p}}{p_0} \nonumber \\
& \quad + \text{\textit{RHS}}_2 - \text{\textit{RHS}}_1 \ .
\label{VarEq}
\end{align}
Now, with proper choice of $\psi$, the variation of the balance equations for the particle number four-flow $N^{\mu}$, energy-momentum tensor $T^{\mu\nu}$ and entropy four-flow $S^{\mu}$ are calculable.

First, for
\begin{align}
& \psi = c, \quad \quad 
 N^{\mu} = c \int p^{\mu} f \sqrt{g} \dfrac{\dd[3]p}{p_0} \nonumber \\
& \bar{\psi} = c , \quad \quad  \bar{N}^{\mu} = c \int p^{\mu} \bar{f} \sqrt{g} \dfrac{\dd[3]p}{p_0}
\end{align}
due to \eqref{VarEq} with the reasonable assumption that $\text{\textit{RHS}}_1 = \text{\textit{RHS}}_2$ \footnote{Note that this assumption means that we are dealing with the changes originated from the curved spacetime and not the effects caused by the change of chemical properties of the system.}, we get
\begin{equation}
\Delta(\nabla_{\mu}N^{\mu}) \equiv \nabla_{\mu}\bar{N}^{\mu} -\nabla_{\mu}N^{\mu}  = 0 \ .
\end{equation}

Second, consider the case for
\begin{align}
& \psi^{\nu} = cp^{\nu}, \quad \quad 
 T^{\mu\nu} = c \int p^{\mu} p^{\nu} f \sqrt{g} \dfrac{\dd[3]p}{p_0} \nonumber \\
& \bar{\psi}^{\nu} = cp^{\nu} , \quad \quad  \bar{T}^{\mu\nu} = c \int p^{\mu} p^{\nu} \bar{f} \sqrt{g} \dfrac{\dd[3]p}{p_0}
\end{align} 

Note that for this case where $\psi$ is 4-vector rather than a scalar, it is straightforward to extend equation \eqref{BEq.p1} to
\begin{equation}
\left[ \int \psi^{\nu} p^{\mu} f \sqrt{g} \dfrac{\dd[3]{p}}{p_0}  \right]_{;\mu} = \int p^{\mu} f (\partial_{\mu} \psi^{\nu}) \sqrt{g} \dfrac{\dd[3]p}{p_0} - \int \pdv{\psi^{\nu}}{p_i} \Gamma^{i}_{\mu\sigma}(X) p^{\mu} p^{\sigma} f \sqrt{g} \dfrac{\dd[3]p}{p_0} + \text{\textit{RHS}}_1 \ .
\end{equation}
Therefore with our assumptions, equation \eqref{VarEq} would lead to
\begin{align}
\left[ \int \psi^{\nu} p^{\mu} \bar{f} \sqrt{g} \dfrac{\dd[3]{p}}{p_0}  \right]_{;\mu} & - \left[ \int \psi^{\nu} p^{\mu} f \sqrt{g} \dfrac{\dd[3]{p}}{p_0}  \right]_{;\mu}
 = - \int p^{\mu} \left( \bar{f} \partial_{\mu} \bar{\psi}^{\nu} - f \partial_{\mu} \psi^{\nu} \right) \sqrt{g} \dfrac{\dd[3]p}{p_0} \nonumber \\
& \quad + \int \left( \pdv{\bar{\psi}^{\nu}}{p_i} \bar{\Gamma}^{i}_{\mu\sigma}(X + \Delta) \bar{f} - \pdv{\psi^{\nu}}{p_i} \Gamma^{i}_{\mu\sigma}(X) f  \right) p^{\mu} p^{\sigma} \sqrt{g} \dfrac{\dd[3]{p}}{p_0} \nonumber \\
& \quad + \text{\textit{RHS}}_2 - \text{\textit{RHS}}_1
\label{ExVarEq}
\end{align}

which leads to
\begin{align}
\Delta(\nabla_{\mu}T^{\mu\nu}) \equiv \nabla_{\mu}\bar{T}^{\mu\nu} -\nabla_{\mu}T^{\mu\nu} & = 0 
\end{align}

Finally, for
\begin{align}
& \psi = - ck_B \ln{\left(\dfrac{f h^3}{e g_s}\right)}, \quad \quad 
S^{\mu} = - ck_B  \int   p^{\mu} \ln{\left(\dfrac{f h^3}{e g_s}\right)}   f \sqrt{g} \dfrac{ \dd[3]{p} }{p_0}  \nonumber \\
& \bar{\psi} = - ck_B \ln{\left(\dfrac{\bar{f} h^3}{e g_s}\right)} , \quad \quad \bar{S}^{\mu} = - ck_B  \int  p^{\mu}  \ln{\left(\dfrac{\bar{f} h^3}{e g_s}\right)} \bar{f}  \sqrt{g} \dfrac{\dd[3]{p}}{p_0}  
\end{align}
we get
\begin{align}
\Delta (\nabla_{\mu} S^{\mu}) & = -\underbrace{\int p^{\mu} \left(\bar{f} \partial_{\mu} \bar{\psi} - f \partial_{\mu} \psi  \right) \sqrt{g} \dfrac{\dd[3]{p}}{p_0}}_{I} \nonumber \\
& + \underbrace{\int \left(  \pdv{\bar{\psi}}{p_i}  \bar{\Gamma}^{i}_{\mu\nu}(X + \Delta) \bar{f} - \pdv{\psi}{p_i} \Gamma^{i}_{\mu\nu} (X) f \right) \sqrt{g} \dfrac{\dd[3]{p}}{p_0}}_{II}  \nonumber \\
& + \text{\textit{RHS}}_2 - \text{\textit{RHS}}_1 \ .
\label{Var. Ent}
\end{align}
Setting $\bar{f} = f+ \delta f$ we have
\begin{align}
\bar{\psi} & = -ck_B \ln{(\dfrac{\bar{f} h^3}{e g_s})} \nonumber \\
&  = -ck_B  \left[  \ln{(\dfrac{f h^3}{e g_s})} + \dfrac{\delta f}{f} + \mathcal{O}((\delta f)^2) \right] \nonumber \\
& \simeq \psi - ck_B \dfrac{\delta f}{f}
\end{align}
and thus the term $I$ in \eqref{Var. Ent} simplifies to 
\begin{align}
I: & \int p^{\mu} \left(\bar{f} \partial_{\mu} \bar{\psi} - f \partial_{\mu} \psi  \right) \sqrt{g} \dfrac{\dd[3]{p}}{p_0} \nonumber \\
& =   \int p^{\mu} \left( (f+ \delta f) \left( \partial_{\mu} \psi -ck_B \partial_{\mu}  \dfrac{\delta f}{f} \right) - f \partial_{\mu} \psi  \right) \sqrt{g} \dfrac{\dd[3]{p}}{p_0}  \nonumber \\
& \simeq  \int p^{\mu} \left( \delta f \partial_{\mu} \psi -ck_Bf \partial_{\mu} \dfrac{\delta f}{f} \right) \sqrt{g} \dfrac{\dd[3]p}{p_0} 
\end{align}
and the term $II$ can be written as
\begin{align}
II: & \int \left(  \pdv{\bar{\psi}}{p_i}  \bar{\Gamma}^{i}_{\mu\nu}(X + \Delta) \bar{f} - \pdv{\psi}{p_i} \Gamma^{i}_{\mu\nu} (X) f \right) \sqrt{g} \dfrac{\dd[3]{p}}{p_0} \nonumber \\
& =  \int \left( \pdv{(\psi - ck_B \dfrac{\delta f}{f})}{p_i}
\left( \Gamma^i_{\mu\nu}(X)  + \Delta^{\alpha} \partial_{\alpha} \Gamma^i_{\mu\nu} \right)
 (f+\delta f)  - \pdv{\psi}{p_i} \Gamma^i_{\mu\nu} f  \right)  \sqrt{g} \dfrac{\dd[3]{p}}{p_0} \nonumber \\
&  =  \int \left( \pdv{\psi}{p_i} \Gamma^i_{\mu\nu}(X) \delta f + \pdv{\psi}{p_i} \Delta^{\alpha} \left(\partial_{\alpha}  \Gamma^i_{\mu\nu}(X) \right) f -ck_B \pdv{}{p_i}\left( \dfrac{\delta f}{f} \right) \Gamma^i_{\mu\nu}(X) f  \right) p^{\mu} p^{\nu} \sqrt{g} \dfrac{\dd[3]{p}}{p_0} \ .
\end{align}
As a result of the assumption of setting $\text{\textit{RHS}}_1 = \text{\textit{RHS}}_2 $, and using the above relations one gets
\begin{multline}
\Delta (\nabla_{\mu} S^{\mu})  = \int \left[ -p^{\mu} \left(\partial_{\mu}\psi \right) \delta f + ck_B \left( \partial_{\mu} \dfrac{\delta f}{f} \right) f p^{\mu}  
+   \left(\partial_{p_i}\psi \right) \delta f \ \Gamma^i_{\mu\nu}(X) p^{\mu} p^{\nu} \right. \\
\left. + \left( \partial_{p_i} \psi \right) f p^{\mu} p^{\nu} \Delta^{\alpha} \partial_{\alpha}  \Gamma^i_{\mu\nu}(x)   - ck_B  f  \partial_{p_i} \left( \dfrac{\delta f}{f}\right)  p^{\mu} p^{\nu} \Gamma^i_{\mu\nu}(X)   \right] \sqrt{g} \dfrac{\dd[3]{p}}{p_0} 
\end{multline}
and since $\psi = -ck_B \ln{\left(\dfrac{f h^3}{e g_s}\right)}$,
\begin{align}
&\Delta (\nabla_{\mu} S^{\mu})  =   \
\begin{multlined}[t]
 ck_B \int \left[  \dfrac{\partial_{\mu} f}{f} \delta f  p^{\mu} +f \partial_{\mu} \left( \dfrac{\delta f}{f} \right)  p^{\mu} -  \dfrac{\partial_{p_i} f}{f} \delta f \Gamma^i_{\mu\nu}(X) p^{\mu} p^{\nu} \right.  \\
 \left. -   f \partial_{p_i}\left(\dfrac{\delta f}{f} \right) \Gamma^i_{\mu\nu}(X) p^{\mu} p^{\nu}  -  \dfrac{\partial_{p_i} f}{f} f  p^{\mu} p^{\nu} \Delta^{\alpha} \partial_{\alpha}  \Gamma^i_{\mu\nu}(X)  \right] \sqrt{g} \dfrac{\dd[3]{p}}{p_0} \end{multlined} \nonumber \\
 & = ck_B \int \left[ \partial_{\mu} \left(\dfrac{\delta f}{f} f\right) p^{\mu} - \partial_{p_i}  \left(\dfrac{\delta f}{f} f\right) \Gamma^i_{\mu\nu}(X) p^{\mu} p^{\nu} -  \left( \partial_{p_i} f \right)  p^{\mu} p^{\nu} \Delta^{\alpha} \partial_{\alpha}  \Gamma^i_{\mu\nu}(X)  \right] \sqrt{g} \dfrac{\dd[3]{p}}{p_0} \nonumber \\
& = ck_B \int \Bigl[ \underbrace{ \partial_{\mu} \left(\delta f \right) p^{\mu} - \partial_{p_i}  \left(\delta f \right) \Gamma^i_{\mu\nu}(X) p^{\mu} p^{\nu} }_{\tilde{Q}(\delta f)} -  \left( \partial_{p_i} f \right)  p^{\mu} p^{\nu} \Delta^{\alpha} \partial_{\alpha}  \Gamma^i_{\mu\nu}(X)  \Bigr] \sqrt{g} \dfrac{\dd[3]{p}}{p_0} \ .
\end{align}
The first two terms in the last bracket (i.e. $\tilde{Q}(\delta f)$) are  of the second order in $\delta f$ and can be ignored. Thus,
\begin{align}
\Delta (\nabla_{\mu} S^{\mu}) & = - ck_B \Delta^{\alpha} \partial_{\alpha}  \Gamma^i_{\mu\nu}(X) \int   \left( \partial_{p_i} f \right)  p^{\mu} p^{\nu}   \sqrt{g} \dfrac{\dd[3]{p}}{p_0} \nonumber \\
& =   2 ck_B \Delta^{\alpha} \partial_{\alpha}  \Gamma^i_{\mu\nu}(X)  \delta^{\mu}_i  \int    p^{\nu}   \sqrt{g} \dfrac{\dd[3]{p}}{p_0} \nonumber \\
& =  2 k_B \Delta^{\alpha} \partial_{\alpha}  \Gamma^i_{i\nu}(X)  N^{\nu} \ .
\label{xyz}
\end{align}

To see what this implies, let us define the entropy as
\begin{equation}
\varsigma = \int \dd[3]{x} \sqrt{g} S^0 \ .
\end{equation}
Since $\nabla_{\mu} S^{\mu} = \dfrac{1}{\sqrt{g}} \partial_{\mu} \left(\sqrt{g} S^{\mu} \right) $, and setting $\nabla_{\mu} S^{\mu} = \cal{Q}$, we have
\begin{equation}
\int \dd[3]{x} \partial_{\mu} \left(\sqrt{g} S^{\mu} \right)  = \int \dd[3]{x} \sqrt{g} \cal{Q}
\end{equation}
which leads to 
\begin{equation}
\dv{t} \varsigma = \int \dd[3]{x} \sqrt{g} \cal{Q} \ .
\end{equation}
Using equation (\ref{xyz}), the relation between entropy rate at points $P_1$ and $P_2$ is given by
\begin{equation}
\dot{\bar{\varsigma}} = \dot{\varsigma} -  2 k_B \int \dd[3]{x} \sqrt{g} \Delta^{\alpha} \partial_{\alpha} \Gamma^{i}_{i \nu}(X) N^{\nu}
\end{equation}
where a dot represents a time derivative.

To make things simpler we can use the Riemann normal coordinate system for which we have $\partial_{\nu} \Gamma^{\mu}_{\alpha \beta} = -\dfrac{1}{3} \left(R^{\mu}_{\ \alpha\beta\nu} + R^{\mu}_{\ \beta\alpha\nu}   \right)$, and thus 
\begin{equation}
\Delta\dot{\varsigma} \equiv \dv{t} \bar{\varsigma} - \dv{t} \varsigma = -\dfrac{2k_B}{3} \left( R^{i}_{\ i\nu\alpha} + R^{i}_{\ \nu i\alpha} \right) \int \dd[3]{x} \sqrt{g} \Delta^{\alpha} N^{\nu} \ .
\label{3rd m-balance eq.}
\end{equation} 
Choosing such a coordinate system means that the points $P_1$ and $P_2$ are two neighboring points in the local (freely falling) frame.

{\em A direct result of equation (\ref{3rd m-balance eq.}) is that if the box travels on a closed loop (with velocity defined by $\Delta^\alpha=U^\alpha \delta t$), the thermodynamics does not restore its initial state.} This shows that such a system remembers its background gravitational history.

Let us summarize what we got up to now. Although the balance equations for the particle number four--flow and the energy--momentum tensor of the fluid in the box would not experience any changes under transportation from point $P_1$ to point $P_2$, the entropy rate will do. From \eqref{3rd m-balance eq.} it is clear that this change is proportional to the Riemann curvature tensor, which means that transporting a small box filled by a fluid, causes the entropy rate to change by a term proportional to the Riemann tensor. Thus, any deviation in the trajectory of the box, whether caused by placing the box in a curved spacetime or by the passage of a gravitational wave pulse, is expected to result in a change in the rate of entropy. This means that the thermodynamics of the box does not return to its initial state, and thus exhibits some hysteresis behavior. This is what we will show in the following.

We have to emphasis again that such an effect is classical in nature and has nothing to do with the quantum mechanical phenomena like particle creation and annihilation caused by gravity.

\section{Gravitational hysteresis}
\label{SecIII}
To investigate the possible consequences of what we derived in the previous section, let us consider the fluid box within various curved backgrounds. The list includes Schwarzschild black hole, Friedmann--Lema\^itre--Robertson--Walker (FLRW) cosmological solution,  and the flat spacetime perturbed by a gravitational wave pulse. In what follows we set $c=1$ to simplify the relations.
\subsection{Schwarzschild black hole spacetime}
First, let us to investigate a small box filled by a fluid located near a Schwarzschild black hole
\begin{equation}
\dd{s}^2 = \left( 1- \frac{2GM}{r} \right) \dd{t}^2 - \left( 1- \frac{2GM}{r} \right)^{-1} \dd{r}^2 -r^2 \dd \Omega \ .
\label{Sch-metric}
\end{equation}
Consider that the box initially is at rest and falling radially, so that (initially) $u^{\mu} = (1,\vb{0})$, and $ N^{\mu} = \rho u^{\mu}=(\rho ,\vb{0})$. We get
\begin{equation}
\Delta\dot{\varsigma} = -\dfrac{2 k_B}{3} \left( R^{1}_{\ 1 0 \alpha} +  R^{1}_{\ 0 1 \alpha} \right) \Delta^{\alpha} \int \dd[3]{x} \sqrt{g} N^0  \ .
\end{equation}
Using the non--vanishing components of the Riemann curvature tensor for the Schwarzschild metric \eqref{Sch-metric}, we would have
\begin{equation}
\Delta\dot{\varsigma} = -\dfrac{2 k_B}{3}  R^{1}_{\ 0 1 0}  \Delta^{0} \rho  \int \dd[3]{x} \sqrt{g} = -\dfrac{8}{9} k_B \dfrac{2m(r-2m)}{r} \Delta^0 \rho
\end{equation}

Integrating $\Delta\dot{\varsigma}$, one may conclude that putting the box outside/inside the horizon would cause the entropy of the fluid to continuously decrease/increase. This is certainly not a physical result. The missed element of the argumentation is the box itself. To maintain the shape of the box in the presence of the gravitational field, an external force (such as tension in the walls) is required. Including the effect of such a force would prevent the entropy from diverging.

But it is quite obvious that the footprint of the gravitational field in the entropy rate would be present and the box remembers the effect of the gravitational field in its thermodynamics. 
It has also needed to note that this kind of entropy change is classical in nature and has nothing to do with Hawking radiation.

On the other hand if the box is in motion on a geodesic around the black hole,  the particle number current would be
\begin{equation}
N^{\mu} = \rho  \left( (1-\dfrac{2GM}{r})^{-1} , \dv{r}{\lambda}, \dfrac{\ell}{r^2}, 0  \right)
\end{equation}
where $\lambda$ is the affine parameter of the geodesic and $\ell$ is the angular momentum of the particle. 
As a result we have
\begin{align}
\Delta\dot{\varsigma}= -\dfrac{2 k_B}{3} \rho  &  \left\{ \left( R^{2}_{\ 121} + R^{3}_{\ 131} +R^{4}_{\ 141} \right) \Delta^0 \int \dot{u}^0 \sqrt{g} \dd[3]x \right.  \nonumber \\
& \left. + \left( R^{1}_{\ 212} + R^{3}_{\ 232} +R^{4}_{\ 242} \right) \Delta^1 \int \dot{u}^1 \sqrt{g} \dd[3]x \right.  \nonumber \\
& \left. + \left( R^{1}_{\ 313} + R^{2}_{\ 323} +R^{4}_{\ 343} \right) \Delta^2 \int \dot{u}^2 \sqrt{g} \dd[3]x  \right\}
\end{align}
It is straightforward to show that this vanishes because of the symmetries of the Riemann tensor.

\subsection{FLRW cosmological spacetime}
As another example consider that our box is placed in a cosmological background. For a co--moving box in a spatially flat cosmological FLRW background
\begin{equation}
\dd{s}^2 = \dd{t}^2 - a^2 \left( \dd{r}^2 + r^2 \dd{\Omega} \right)
\end{equation}
where the $a \equiv a(t)$ is the scale factor, the $\Delta\dot{\varsigma}$ can be obtained in a similar manner as:
\begin{equation}
\Delta\dot{\varsigma} = \dfrac{2 k_B}{3} \dfrac{\ddot{a}}{a} \Delta^0 \rho  \int \dd[3]{x} \sqrt{g} = \dfrac{2 k_B}{3}\rho {\cal V}a^2\ddot{a}\Delta^0
\end{equation}
where ${\cal V}$ is the box volume. Setting $\Delta^0=\delta t$ and integrating $\Delta\dot{\varsigma}$, we get
\begin{equation}
\varsigma(t)=\varsigma(t_0)+\dfrac{2 k_B}{3} {\cal V} \int_{t_0}^t\!dt_1\int_{t_0}^{t_1}\!dt_2\ \rho\ a^2\ddot{a}
\end{equation}
Therefore even if the cosmological background comes back to its original state (like the oscillating model of the universe), the entropy of the fluid in the box does not come back to its original value. This is a hysteresis effect.

\subsection{Gravitational wave}
As a final example, let's investigate how the entropy rate of the box will change by the passage of a gravitational wave pulse. Consider a pulse propagating in the $\vu{n}$ direction  with polarization amplitudes  $h_+$ and $h_{\times}$ in the flat spacetime. The metric would be
\begin{equation}
g_{\mu\nu}^{\text{(Cartesian)}}=\eta_{\mu\nu}+h_{\mu\nu} =
\begin{pmatrix}
-1 & 0 & 0 & 0 \\
0 & 1+h_+ ~ f(t,r) & h_{\times} ~ f(t,r)  & 0 \\
0 & h_{\times} ~ f(t,r) & 1 -h_+ ~ f(t,r) & 0 \\
0 & 0 & 0 & 1
\end{pmatrix}
\end{equation}
in the Cartesian coordinates $(t,x,y,z)$. Here $f(t,r) \propto e^{-i(\omega t- \vb{k} \vdot \vb{r})}$ is a localized function of the retarded time and spatial distance $r \equiv \sqrt{x^2+y^2+z^2} $ from the source. Without loss of generality we consider that the propagation is along the $z$-axis, and thus, $f(t,r) \propto e^{-i(\omega t- k z)}$.
Up to the leading quadrupolar order, the polarization amplitudes $ h_+ $ and $h_{\times}$  depend on the second time derivative of the second mass moments ($M_{ij}$) of the source as,
\begin{align}
h_+ & = \dfrac{1}{r} G\left(\ddot{M}_{xx} - \ddot{M}_{yy} \right) \nonumber \\
h_{\times} & = \dfrac{2}{r} G \ddot{M}_{xy} \ .
\end{align}
To simplify the model, we assume that polarization amplitudes are constant during the passage of pulse through the box and use
\begin{align}
h_+ & = \dfrac{H_+}{r} \nonumber \\
h_{\times} & = \dfrac{H_{\times}}{r}
\end{align}
instead, where $H_+$ and $H_-$ are constants due to our consideration.
This assumption is quite reasonable if, for example, we are dealing with a wave pulse whose duration is longer than the period of the wave. In such a case we are allowed to use the time average of the second time derivative of the second mass moments, i.e. $\overline{\ddot{M}}_{ij}$ instead of using $\ddot{M}_{ij}$, as an approximation (see \cite{Moti:2023tey}).

For a box at rest, $u^{\mu} = (1,\vb{0})$, and $ N^{\mu} = \rho u^{\mu}=(\rho ,\vb{0})$, and therefore using the nonvanishing components of the Riemann tensor, the entropy rate change would be
\begin{align}
\Delta\dot{\varsigma} =  -\dfrac{2k_B}{3} &\left(   R^{i}_{\ i 0 \alpha} + R^{i}_{\ 0 i\alpha} \right) \Delta^{\alpha} \int \dd[3]{x} \sqrt{g}  N^{0} \nonumber \\
 = -\dfrac{2k_B}{3} & \left( \left( R^{1}_{\ 010} +R^{2}_{\ 020} \right) \Delta^0 + \left( R^1_{\ 101}+R^{2}_{\ 201} +R^{2}_{021} \right) \Delta^1 \right.  \nonumber \\
& \left. + \left(R^{1}_{\ 102}+R^{2}_{\ 202}+R^{1}_{012} \right) \Delta^2  + \left(R^1_{\ 013}+R^{2}_{\ 0 23} \right) \Delta^4 \right) \int  N ^0 \sqrt{g} \dd[3]x \ .
\end{align} 
For the simple choice of $\Delta = (0,0,0,\Delta^z)$, we arrive at
\begin{equation}
\Delta\dot{\varsigma} =  - \dfrac{2k_B}{3}  \omega  \left(H_+^2 + H_{\times}^2 \right) 
\dfrac{ \left(  H_+^2 + H_{\times}^2  - 3 r^2 e^{2i(\omega t - kz) }  \right) \left( i z + kr^2 \right)}{ 2 r^2 \left(    H_+^2 + H_{\times}^2  - e^{2i(\omega t-kz)} r^2 \right)^2 } 
\Delta^z \rho {\cal V}
\end{equation}

This relation can be integrated over time and just like what we obtained in the previous subsection it shows that the entropy remembers the footprint of the passage of the gravitational wave pulse, and that this footprint remains in the thermodynamics of the fluid in the box. It is important to note that due to our assumption of using the time average of $\ddot{M}_{ij}$, the quantity $\Delta\dot{\varsigma}$ does not refer to the instantaneous change in the entropy rate within the box. Instead, it reflects the overall magnitude of the changes rather than the instantaneous periodic behavior. This is a rough but good estimate of the entropy rate. If our assumption is not valid, what would be obtained for the entropy rate would have an additional oscillatory time dependence with the frequency of the wave pulse.

\section{Conclusion}
The main result obtained here is this:

\begin{center}
\parbox{12cm}{\vbox{
{\em The difference between the entropy rate of a fluid in a small box, when located at some point in the spacetime or at some neighbouring point is proportional to the Riemann tensor at that point.}}}
\end{center}

The two neighbouring points could be separated in a space-like or time-like manner. One special case is to consider a specified background (like a black hole or a cosmological background) and compare the entropy rate either at two spatially separated points or at a spatial point at two times. We saw that this leads to a change in entropy rate proportional to the Riemann tensor. 

In fact, if one assumes that the box is transported from point $P_1$ to point $P_2$ by the velocity 4-vector $U^\alpha=\Delta^\alpha/\Delta^0$, equation (\ref{3rd m-balance eq.}) can be rewritten as
\begin{equation}
\ddot{\varsigma} = -\dfrac{2k_B}{3} \left( R^{i}_{\ i\nu\alpha} + R^{i}_{\ \nu i\alpha} \right) \int \dd[3]{x} \sqrt{g} U^{\alpha} N^{\nu}
\end{equation}
which upon integration shows that the entropy would not come back to its initial value if the box makes a trip on a closed loop. {\em This shows that the entropy will memorize what happens to its gravitational background.}

Another possibility to observe such an effect, is to put a fluid box in the vicinity of the passage of a gravitational wave pulse. We saw that again for such a configuration the system shows hysteresis behavior. 

\appendix
\renewcommand{\thesection}{Appendix}
\renewcommand{\theequation}{\Alph{section}.\arabic{equation}}
\setcounter{equation}{0}

\section{Review of Kinetic Theory in Curved Backgrounds} \label{App. A}
Here, we briefly review the Kinetic theory in curved spacetime.
The Boltzmann equation in curved spacetime \cite{Cercignani:2002} 
\begin{equation}
p^{\mu} \pdv{f}{x^{\mu}} - \Gamma^{\lambda}_{\mu\nu}(X) p^{\mu} p^{\nu} \pdv{f}{p^{\lambda}} = Q(f,\hat{f},F,\Omega,p)
\label{BEqCSP}
\end{equation}
characterizes the evolution of the distribution function ($f(x,p)$) of particles of a fluid.
The $\Omega$ is the solid angle in the cross section of the binary collision between the particles of the fluid and $F$ is the invariant flux, defined as {$F= \sqrt{(p^{\mu}_* p_{\mu})^2 - m^4 c^4} = \sqrt{(g_{\mu\nu}p^{\mu}_* p^{\nu})^2 - m^4 c^4}$.
Any starred quantity refers to a specific particle that collided with a non-starred particle.
And the $\sigma$ is the invariant differential cross-section. 
Although in general, there is no invariant constraint on the form of the distribution function $f$ and thus it may change to $\hat{f}$ due to the collision, but for the 1-particle distribution function and using the molecular chaos hypothesis ({\it Stosszahlansatz}) in the case of local equilibrium the distribution function before and after the collision are the same.  Therefore we can set $\hat{f}=f$ in the collision term (see \cite{Cercignani:2002}).
For simplicity, we show $Q(f,\hat{f},F,\Omega,p)=Q(f,f,F,\Omega,p)$ as $Q$ in what follows.

In order to describe macroscopic properties of the fluid, one can use the method of moments. By definition, the nth--moment of the one-particle distribution function is defined as  \cite{Cercignani:2002}
\begin{equation}
T^{\alpha\beta\cdots\gamma\delta} = c \int p^{\alpha} p^{\beta} \cdots p^{\gamma} p^{\delta} f \dfrac{\dd[3]p}{p_0}
\end{equation}
These moments provide us with a deeper understanding of the physical treatments of the system without requiring the direct solution of the Boltzmann equation. Among these moments, the lower-ranked ones are particularly valuable. The first moment is particle number four-flow
\begin{equation}
N^{\alpha} = c \int p^{\alpha} f \dfrac{\dd[3]{p}}{p_0}  
\end{equation}
and the second one is the energy-momentum tensor
\begin{equation}
T^{\alpha\beta} = c \int p^{\alpha} p^{\beta} f \dfrac{\dd[3]{p}}{p_0} 
\end{equation}
Using the transfer equation for a single relativistic fluid 
\begin{equation}
\pdv{}{x^{\alpha}} \int \psi p^{\alpha} f \dfrac{\dd[3]p}{p_0} - \int \left[ p^{\alpha} \pdv{\psi}{x^{\alpha}} + m K^{\alpha} \pdv{\psi}{p^{\alpha}}  \right] f \dfrac{\dd[3]p}{p_0} = \mathcal{P}
\end{equation}
where 
\begin{equation}
\mathcal{P} = \dfrac{1}{4} \int \left( \psi + \psi_* - \psi' -\psi'_* \right) \left(f'_* f' - f_* f \right) F \sigma \dd{\Omega} \dfrac{\dd[3]p_*}{p_{*0}} \dfrac{\dd[3]p}{p_0}
\end{equation}
for non-degenrate gas and
\begin{multline}
\mathcal{P} = \dfrac{1}{4} \int \left( \psi + \psi_* - \psi' -\psi'_* \right) \times \\\left[f'_* f' \left(1+ \epsilon \dfrac{fh^3}{g_s} \right) \left(1+ \epsilon \dfrac{f_*h^3}{g_s} \right) - f_* f  \left(1+ \epsilon \dfrac{f'h^3}{g_s} \right) \left(1+ \epsilon \dfrac{f'_*h^3}{g_s} \right)\right] F \sigma \dd{\Omega} \dfrac{\dd[3]p_*}{p_{*0}} \dfrac{\dd[3]p}{p_0}
\end{multline}
for the degenerate one, the balance equation of each of these moments is derived with proper choices of $\psi \equiv \psi(x^{\mu},p^{i})$.

To this end, for a chosen scalar function $\psi$, integrating the Boltzman equation \eqref{BEqCSP} over all values of $ (g/p_0) \dd[3]{p} \dd[4]{x}$, we get
\begin{equation}
\int \psi \left[ p^{\mu} \pdv{f}{x^{\mu}} - \Gamma^{i}_{\mu\nu} p^{\mu} p^{\nu} \pdv{f}{p^{i}} \right] g \dfrac{\dd[3]{p}}{p_0}  \dd[4]{x} = \int Qg \dfrac{\dd[3]{p}}{p_0}   \dd[4]{x}
\label{MainEq}
\end{equation}
The left hand side of the above equation can be simplified as
\begin{align}
\text{LHS} = &  \int \psi  p^{\mu} \pdv{f}{x^{\mu}}  \dfrac{g}{p_0} \dd[3]{p} \dd[4]{x} -  \int \psi \Gamma^{i}_{\mu\nu} p^{\mu} p^{\nu} \pdv{f}{p^{i}} \dfrac{g}{p_0} \dd[3]{p} \dd[4]{x} \nonumber \\
\begin{split} = & 
 \int \pdv{ }{x^{\mu}}\left( \psi  p^{\mu} f  \dfrac{g}{p_0} \right)  \dd[3]{p} \dd[4]{x} -  \int \pdv{\psi}{x^{\mu}}  p^{\mu} f  \dfrac{g}{p_0} \dd[3]{p} \dd[4]{x} -  \cancelto{\text{LT}}{\int \psi  f \pdv{ }{x^{\mu}} \left( p^{\mu}\dfrac{g}{p_0} \right)  \dd[3]{p} \dd[4]{x}}  \\
&  -  \cancelto{0}{\int \pdv{ }{p^{i}}\left( \psi \Gamma^{i}_{\mu\nu} p^{\mu} p^{\nu} f  \dfrac{g}{p_0} \right) \dd[3]{p} \dd[4]{x}} + \int \pdv{\psi}{p^{i}} \Gamma^{i}_{\mu\nu} p^{\mu} p^{\nu} f \dfrac{g}{p_0} \dd[3]{p} \dd[4]{x}  \\
& + \cancelto{\text{LT}}{ \int \psi f  \pdv{ }{p^{i}}\left(  \Gamma^{i}_{\mu\nu} p^{\mu} p^{\nu}  \dfrac{g}{p_0} \right) \dd[3]{p} \dd[4]{x}}
\end{split} 
\end{align}
where the terms denoted by LT are canceled as a result of the Liouville Theorem and thus
\begin{align}
LHS  = & \int \pdv{ }{x^{\mu}}\left( \psi  p^{\mu} f \dfrac{g}{p_0} \right)  \dd[3]{p} \dd[4]{x}-  \int \pdv{\psi}{x^{\mu}}  p^{\mu} f  \dfrac{g}{p_0} \dd[3]{p} \dd[4]{x}  + \int \pdv{\psi}{p^{i}} \Gamma^{i}_{\mu\nu} p^{\mu} p^{\nu} f \dfrac{g}{p_0} \dd[3]{p} \dd[4]{x} \nonumber \\
= &  \int \pdv{ }{x^{\mu}}\left( \psi  p^{\mu} f  \dfrac{\sqrt{g}}{p_0} \right) \sqrt{g} \dd[3]{p} \dd[4]{x} +  \int \psi  p^{\mu} f  \dfrac{g}{p_0} \left(\partial_{\mu} \sqrt{g} \right)  \dd[3]{p} \dd[4]{x} \nonumber \\
&  -  \int \pdv{\psi}{x^{\mu}}  p^{\mu} f  \dfrac{g}{p_0} \dd[3]{p} \dd[4]{x} + \int \pdv{\psi}{p^{i}} \Gamma^{i}_{\mu\nu} p^{\mu} p^{\nu} f \dfrac{g}{p_0} \dd[3]{p} \dd[4]{x}\nonumber \\
 = & \int \nabla_{\mu} \left( \psi  p^{\mu} f  \dfrac{\sqrt{g}}{p_0} \right) \sqrt{g} \dd[3]{p} \dd[4]{x}   -  \int \pdv{\psi}{x^{\mu}}  p^{\mu} f  \dfrac{g}{p_0} \dd[3]{p} \dd[4]{x} + \int \pdv{\psi}{p^{i}} \Gamma^{i}_{\mu\nu} p^{\mu} p^{\nu} f \dfrac{g}{p_0} \dd[3]{p} \dd[4]{x} \nonumber \\
\end{align}
Therefore we get
\begin{equation}
LHS = \int \sqrt{g} \dd[4]{x} \int \left(  \nabla_{\mu} \left( \psi  p^{\mu} f  \dfrac{\sqrt{g}}{p_0} \right)- \pdv{\psi}{x^{\mu}}  p^{\mu} f  \dfrac{\sqrt{g}}{p_0} + \pdv{\psi}{p^{i}} \Gamma^{i}_{\mu\nu} p^{\mu} p^{\nu} f \dfrac{\sqrt{g}}{p_0}  \right)\dd[3]{p}
\label{LHS-App}
\end{equation}
On the other hand, the right hand side of the equation \eqref{MainEq}, is
\begin{equation}
\text{\textit{RHS}} = \int \sqrt{g} \dd[4]{x} \int \tilde{Q} \sqrt{g}  \dfrac{\dd[3]{p} }{p_0}
\label{RHS-App}
\end{equation}
where $\tilde{Q}$ is the same as $Q$ using the symmetry considerations.

Finally, equation \eqref{MainEq} reduces to
\begin{equation}
\int  \nabla_{\mu} \left( \psi  p^{\mu} f  \dfrac{\sqrt{g}}{p_0} \right)  \dd[3]{p} - \int  \pdv{\psi}{x^{\mu}}  p^{\mu} f  \dfrac{\sqrt{g}}{p_0}  \dd[3]{p} + \int  \pdv{\psi}{p^{i}} \Gamma^{i}_{\mu\nu} p^{\mu} p^{\nu} f \dfrac{\sqrt{g}}{p_0}  \dd[3]{p} =  \int \tilde{Q}(f,f) \sqrt{g} \dfrac{\dd[3]{p} }{p_0} 
\label{Entropy-MainEq}
\end{equation}

The proper choice for $\psi$ to derive the balance equations for the essential moments \eqref{EsMom} are
\begin{equation}
\psi =c, \quad \quad \psi^{\nu} = c p^{\nu},  \quad \quad \psi = - ck_B \ln{(\dfrac{f h^3}{e g_s})}
\end{equation}
respectively, leading to the corresponding equations
\begin{equation}
N^{\mu}_{\ ;\mu} = 0 , \quad \quad T^{\mu\nu}_{\ \ ;\mu}  = 0 , \quad \quad S^{\mu}_{\ ;\mu} \geq 0
\end{equation}

The derivation of the first two is simple. To obtain the relation for entropy, we substitute $\psi = -ck_B \ln{(\dfrac{f h^3}{e g_s})}$ into equation \eqref{Entropy-MainEq} and rewrite it as
\begin{multline}
-ck_B  \int  \nabla_{\mu} \left( \ln{(\dfrac{f h^3}{e g_s})} p^{\mu} f  \dfrac{\sqrt{g}}{p_0} \right)  \dd[3]{p} + ck_B \int  \partial_{\mu} \ln{(\dfrac{f h^3}{e g_s})} p^{\mu} f  \dfrac{\sqrt{g}}{p_0}  \dd[3]{p} \\
 -ck_B \int  \pdv{p^{i}} (\ln{(\dfrac{f h^3}{e g_s})})  \Gamma^{i}_{\mu\nu} p^{\mu} p^{\nu} f \dfrac{\sqrt{g}}{p_0}  \dd[3]{p} =  \int \tilde{Q} \dfrac{\sqrt{g} }{p_0} \dd[3]{p}
\end{multline}
Then,
\begin{align}
S^{\mu}_{;\mu} & = -ck_B \left[ \int  \partial_{\mu} \ln{(\dfrac{f h^3}{e g_s})} p^{\mu} f  \sqrt{g} \dfrac{\dd[3]{p} }{p_0} -  \int  \pdv{p^{i}} (\ln{(\dfrac{f h^3}{e g_s})})  \Gamma^{i}_{\mu\nu} p^{\mu} p^{\nu} f \sqrt{g}\dfrac{\dd[3]{p} }{p_0}  \right] +  \int \tilde{Q}(f,f) \sqrt{g}\dfrac{\dd[3]{p}}{p_0}  \nonumber \\
& = -ck_B \left[ \int p^{\mu} f \dfrac{\partial_{\mu} f}{f} \sqrt{g} \dfrac{\dd[3]p}{p_0} -\int \dfrac{\pdv{f}{p_i}}{f} \Gamma^{i}_{\mu\nu} p^{\mu} p^{\nu} f \sqrt{g} \dfrac{\dd[3]{p}}{p_0} \right]  \nonumber \\
& \quad  \underbrace{ - \dfrac{ck_B}{4} \int \ln{(\dfrac{f f_*}{f' f'_*})} \left(f' f'_* -  f f_* \right) F \sigma \dd{\Omega} \sqrt{g} \dfrac{\dd[3]{p_*}}{p_{*0}} \sqrt{g} \dfrac{\dd[3]{p}}{p_0} }_{\text{Last Term}} 
\end{align}
with the final term being derived from the internal interactions of the system, as elaborated in detail in \cite{Cercignani:2002}.

The integrand of the last term is negative, since $(1-X) \ln X$ is negative for any positive $X$ including $X \equiv X_f = \dfrac{f f_*}{f' f'_*}$, which means that this term is positive and on using the Boltzmann equation, one arrives at $S^{\mu}_{\ ;\mu} \geq 0$.


\begin{thebibliography}{10}

\bibitem{Braginsky:1987kwo}
V.B. Braginsky and K.S. Thorne, Gravitational-wave bursts with memory and experimental prospects, \href{https://doi.org/10.1038/327123a0}{Nature \textbf{327}, 123  (1987)}.


\bibitem{Favata:2009ii}
M. Favata, Nonlinear gravitational-wave memory from binary black hole mergers, 
\href{https://doi.org/10.1088/0004-637X/696/2/L159}{Astrophys. J. Lett. \textbf{696} (2009)}.

\bibitem{Zeldovich:1974gvh}
Y.B. Zel'dovich and A.G. Polnarev, Radiation of gravitational waves by a cluster of superdense stars, \href{https://ui.adsabs.harvard.edu/abs/1974AZh....51...30Z}{Sov. Astron. \textbf{18}, 17 (1974)}.

\bibitem{Smarr:1977fy}
L. Smarr, Gravitational radiation from distant encounters and from headon collisions of black holes: The zero frequency limit, \href{https://doi.org/10.1103/PhysRevD.15.2069}{Phys. Rev. D \textbf{15}, 2069 (1977)}.

\bibitem{Kovacs:1978eu}
S.J. Kovacs and K.S. Thorne, The generation of gravitational waves. \Romannum{4}. Bremsstrahlung, \href{https://doi.org/10.1086/156350}{Astrophys. J. \textbf{224}, 62 (1978)}.

\bibitem{Braginsky:1985vlg}
V.B. Braginsky and L.P. Grishchuk, Kinematic resonance and memory effect in free mass gravitational antennas, \href{https://ui.adsabs.harvard.edu/abs/1985ZhETF..89..744B/abstract}{Sov. Phys. JETP \textbf{62}, 427 (1985)}.

\bibitem{Favata:2008ti}
M. Favata, Gravitational-wave memory revisited: Memory from the merger and recoil of binary black holes, \href{https://doi.org/10.1088/1742-6596/154/1/012043}{J. Phys. Conf. Ser. \textbf{154}, 012043 (2009)}.

\bibitem{Favata:2008yd}
M. Favata, Post-Newtonian corrections to the gravitational-wave memory for quasi-circular, inspiralling compact binaries, \href{https://doi.org/10.1103/PhysRevD.80.024002}{Phys. Rev. D \textbf{80}, 024002 (2009)}.

\bibitem{Christodoulou:1991cr}
D. Christodoulou, Nonlinear nature of gravitation and gravitational wave experiments, \href{https://doi.org/10.1103/PhysRevLett.67.1486}{Phys. Rev. Lett. \textbf{67}, 1486 (1991)}.

\bibitem{Blanchet:1992br}
L. Blanchet and T. Damour, Hereditary effects in gravitational radiation, \href{https://doi.org/10.1103/PhysRevD.46.4304}{Phys. Rev. D, \textbf{46}, 4304 (1992)}.

\bibitem{Thorne:1992sdb}
K. S. Thorne, Gravitational-wave bursts with memory: The Christodoulou effect, \href{https://doi.org/10.1103/PhysRevD.45.520}{ Phys. Rev. D \textbf{45}(2), 520 (1992)}.

\bibitem{Favata:2010zu}
M. Favata, The gravitational-wave memory effect, \href{https://doi.org/10.1088/0264-9381/27/8/084036}{Class. Quant. Grav. \textbf{27}, 084036 (2010)}.

\bibitem{Zhang:2017geq}
P.M. Zhang, C.Duval, G.W. Gibbons, and P.A. Horvathy, Soft gravitons and the memory effect for plane gravitational waves, \href{https://doi.org/10.1103/PhysRevD.96.064013}{Phys. Rev. D, \textbf{96}(6), 064013 (2017)}.

\bibitem{Grishchuk:1989qa}
L.P. Grishchuk and A.G. Polnarev, Gravitational wave pulses with `velocity coded memory', \href{https://ui.adsabs.harvard.edu/abs/1989ZhETF..96.1153G/abstract}{Sov. Phys. JETP \textbf{69}, 653 (1989)}.

\bibitem{Divakarla:2021xrd}
A.K. Divakarla and B.F. Whiting, First-order velocity memory effect from compact binary coalescing sources, \href{https://doi.org/10.1103/PhysRevD.104.064001}{Phys. Rev. D \textbf{104}(6), 064001 (2021).}

\bibitem{Barnich:2009se}
G. Barnich and C. Troessaert, Symmetries of asymptotically flat 4 dimensional spacetimes at null infinity revisited, \href{https://doi.org/10.1103/PhysRevLett.105.111103}{Phys. Rev. Lett. \textbf{105}, 111103 (2010)}.

\bibitem{Pasterski:2015tva}
S. Pasterski, A. Strominger, and A. Zhiboedov, New gravitational memories, \href{https://doi.org/10.1007/JHEP12(2016)053}{JHEP \textbf{12}, 053 (2016)}.

\bibitem{Nichols:2017rqr}
D.A. Nichols, Spin memory effect for compact binaries in the post-Newtonian approximation, \href{https://doi.org/10.1103/PhysRevD.95.084048}{Phys. Rev. D, \textbf{95}(8), 084048 (2017)}.

\bibitem{Seraj:2021rxd}
A. Seraj and B. Oblak, Gyroscopic gravitational memory, \href{https://doi.org/10.1007/JHEP11(2023)057}{JHEP \textbf{11}, 057 (2023)}.

\bibitem{Seraj:2022qyt}
A. Seraj and B. Oblak, Precession caused by gravitational waves,\href{https://doi.org/10.1103/PhysRevLett.129.061101} {Phys. Rev. Lett. \textbf{129}(6), 061101 (2022)}.

\bibitem{Moti:2023tey}
R. Moti and A. Shojai, On the gravitational precession memory effect for an ensemble of gyroscopes, \href{https://doi.org/10.1088/1361-6382/ad1780}{Class. Quant. Grav. \textbf{41}(2), 025011 (2024)}.

\bibitem{Cercignani:2002}
C. Carlo and G. Kremer, \textit{The relativistic Boltzmann equation: Theory and applications}, (Birkhauser Basel, 2002).

\end{thebibliography}
\end{document}